\documentclass[sigconf]{acmart}
\usepackage{graphicx}
\graphicspath{{figures/},{pics/}}
\usepackage{xspace}
\usepackage{algorithmic}
\usepackage{subfigure}
\usepackage{graphicx}
\usepackage{textcomp}
\usepackage{xcolor}
\usepackage{multirow}
\usepackage{tcolorbox}
\usepackage{booktabs}
\usepackage{url}
\AtBeginDocument{%
  }
\AtBeginDocument{%
  }

\newcommand{\etal}{{\em et al.}\xspace}
\newcommand{\ie}{{\em i.e.}\xspace}
\newcommand{\eg}{{\em e.g.}\xspace}

\setcopyright{acmlicensed}
\copyrightyear{2024}
\acmYear{2024}
\acmDOI{XXXXXXX.XXXXXXX}

\acmConference[Conference acronym 'XX]{Make sure to enter the correct
  conference title from your rights confirmation emai}{Oct 27--Nov 1,
  2024}{Sacramento, CA, USA}
\acmISBN{978-1-4503-XXXX-X/24/10}

\acmSubmissionID{658}

\begin{document}

\title{Unlocking the Power of Numbers:\\ Log Compression via Numeric Token Parsing}

\author{Siyu Yu}
\email{gaiusyu6@gmail.com}
\orcid{0000-0003-3793-7684}
\affiliation{%
  \institution{The Chinese University of Hong Kong, Shenzhen (CUHK-Shenzhen)}
  \city{Shenzhen}
  \country{China}
}

\author{Yifan Wu}
\email{yifanwu@pku.edu.cn}
\orcid{0000-0001-5847-3132}
\affiliation{%
  \institution{Peking University}
  \city{Beijing}
  \country{China}}

\author{Ying Li}
\email{li.ying@pku.edu.cn}
\orcid{0000-0002-6278-2357}
\affiliation{%
  \institution{Peking University}
  \city{Beijing}
  \country{China}}

\author{Pinjia He}
\email{hepinjia@cuhk.edu.cn}
\orcid{0000-0003-3377-8129}
\authornote{Pinjia He is the corresponding author of this work. E-mail: hepinjia@cuhk.edu.cn}
\affiliation{%
 \institution{The Chinese University of Hong Kong, Shenzhen (CUHK-Shenzhen)}
  \institution{Shenzhen Research Institute of Big Data}
 \city{Shenzhen}
 \country{China}}

\renewcommand{\shortauthors}{Yu et al.}

\begin{abstract}
Parser-based log compressors have been widely explored in recent years because the explosive growth of log volumes makes the compression performance of general-purpose compressors unsatisfactory.
These parser-based compressors preprocess logs by grouping the logs based on the parsing result and then feed the preprocessed files into a general-purpose compressor. 
However, parser-based compressors have their limitations.
First, the goals of parsing and compression are misaligned, so the inherent characteristics of logs were not fully utilized.
In addition, the performance of parser-based compressors depends on the sample logs and thus it is very unstable.
Moreover, parser-based compressors often incur a long processing time.
To address these limitations, we propose \textit{Denum}, a simple, general log compressor with high compression ratio and speed.
The core insight is that a majority of the tokens in logs are numeric tokens (\ie pure numbers, tokens with only numbers and special characters, and numeric variables) and effective compression of them is critical for log compression.
Specifically, \textit{Denum} contains a \textit{Numeric Token Parsing} module, which extracts all numeric tokens and applies tailored processing methods (\eg store the differences of incremental numbers like timestamps), and a \textit{String Processing} module, which processes the remaining log content without numbers. 
The processed files of the two modules are then fed as input to a general-purpose compressor and it outputs the final compression results. 
\textit{Denum} has been evaluated on 16 log datasets and it achieves an $8.7\%-434.7\%$ higher average compression ratio and $2.6\times$ $-$ $37.7\times$ faster average compression speed (\ie~26.2 MB/S) compared to the baselines. Moreover, integrating \textit{Denum}'s \textit{Numeric Token Parsing} module into existing log compressors can provide a $11.8\%$ improvement in their average compression ratio and achieve $37\%$ faster average compression speed.
\end{abstract}



\received{7 June 2024}
\received[accepted]{6 Aug 2024}

\maketitle

\section{Introduction}
Logs capture important system runtime information \cite{gholamian2021comprehensive,chen2020improving,he2021survey,chen2021survey}, making them a critical data source for maintenance and operation tasks, such as anomaly detection \cite{du2017deeplog}, root cause analysis \cite{lu2017log}, and system state modeling \cite{chow2014mystery}. Typically, logs need to be preserved for a certain period of time for post-mortem analysis. For example, AliCloud requires logs to be stored for a minimum of 180 days.

In recent years, the volume of logs has experienced exponential growth.
In 2016, Feng \etal~\cite{feng2016mlc} reported that their system generated 100 GB logs daily, and in 2019, the volume of logs generated by modern systems per day surged to 2 TB~\cite{liu2019logzip}. 
By 2021, AliCloud collected daily logs at the PB (petabyte) scale~\cite{wei2021feasibility}. 
Thus, as storage costs escalate, effective and efficient log compression becomes necessary. For instance, Google Cloud bills as much as \$465,700 for storing 1 PB of data for one month \cite{googlecost1}. 
A straightforward way to compress log files is to use general-purpose compressors like LZMA \cite{lzmalink}, PPMd \cite{ppmdlink}, gzip \cite{gziplink}, and bzip2 \cite{bzip2link}. These tools are versatile and capable of compressing different kinds of data. Meanwhile, it did not consider the inherent structure of logs, resulting in sub-optimal performance for log data.

To better exploit the inherent structure within logs, researchers have developed various log compressors. These compressors first utilize the structure of logs to transform the raw log files into various types of preprocessed files, after which general-purpose compressors are applied to these preprocessed files. A primary category of log compressors is parser-based log compressors, which utilize log parsers to exploit log structure, such as LogZip \cite{liu2019logzip}, LogReducer \cite{wei2021feasibility}, and LogShrink \cite{li2024logshrink}. These tools first transform semi-structured logs into structured logs using log parsers \cite{he2021survey} like Drain \cite{he2017drain} and then group logs by the same log template for tailored different preprocessing schemes to generate different preprocessed files. 
Besides these parser-based log compressors, there are log compressors that apply unified compression strategies to both templates and variables, such as the LogAchieve \cite{christensen2013adaptive}. 

\begin{figure}[t]
	\centering
		\includegraphics[width=0.95\columnwidth]{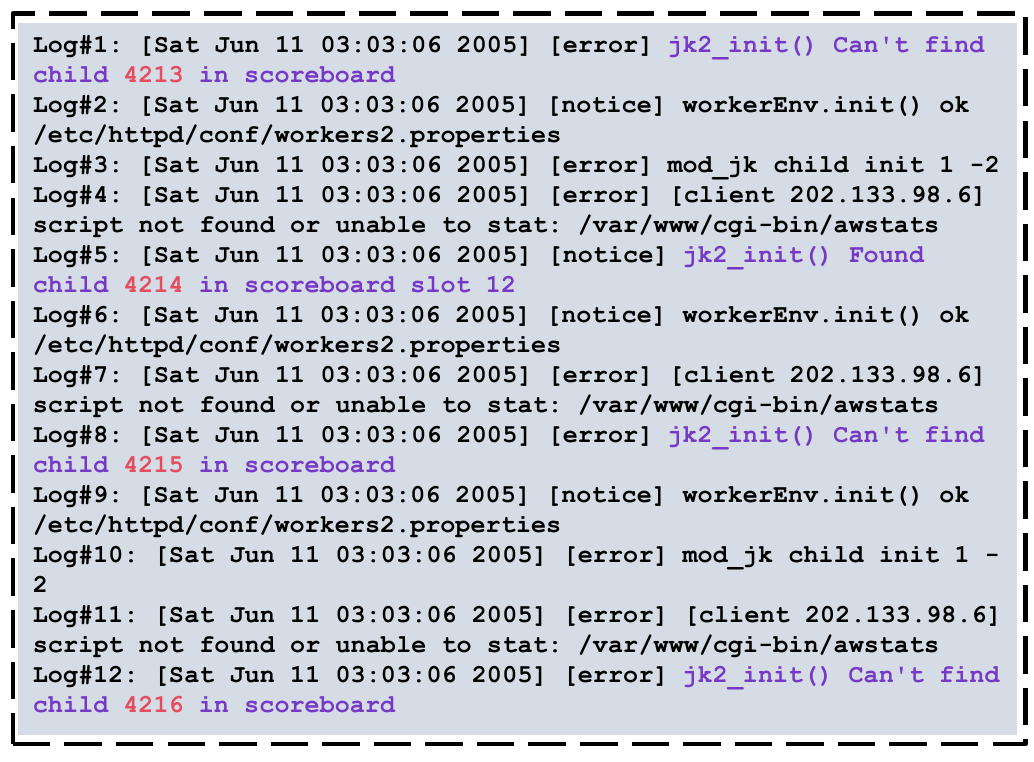}
	  \caption{Increasing numbers across different Templates. The increasing numbers are in red and the templates these numbers belong to are in purple.}\label{crosstemplate}
\end{figure}

Despite the effectiveness of parser-based log compressors, they are unsatisfactory for industry and have the following limitations.
\textit{First}, the goal of log parsing is misaligned with the goal of log compression, so the inherent characteristics of logs (\eg similarity between numeric tokens that span multiple log templates) have not been fully utilized.
For example, in Figure \ref{crosstemplate}, the incremental numbering of "\verb|child|" in $log\#1, log\#5, log\#8, log\#12$ spans two templates: "\verb|Can’t find child <*> in scoreboard|" and "\verb|Found child <*>|\\\verb|in seaboard slog <*>|", incrementing from 4213 to 4215 with a step size of 1. 
Parsing and grouping by templates will separate these incrementing numbers. 
\textit{Second}, the performance of parser-based compressors depends on the sample logs and thus it is very unstable. 
LogShrink~\cite{li2024logshrink} samples the logs and parses the sampled logs to generate log templates, which are further used to match the upcoming logs. 
Its compression ratio can change from 13 to 65 on HealthApp dataset, where a "13" compression ratio is worse than that provided by a general-purpose compressor.
\textit{Third}, parser-based compressors often incur a long processing time. 

We observe that logs contain many numeric tokens by nature, such as timestamps, counts, thread/process IDs, and block numbers. 
Up to 70\% of the preprocessed log files generated by existing log compressors were used to store numeric tokens. 
Thus, we hypothesize that effectively compressing numbers is the key to log compression.
Meanwhile, existing research has overlooked that general-purpose compressors are good at compressing repetitive tokens.
For example, the core of LZMA is the Lempel-Ziv algorithm~\cite{ziv1977universal, ziv1978compression}, which uses a dictionary to store previously occurring elements and can covert repetitive template words into more concise representations. Therefore, removing the parsing step may not significantly reduce the compression ratio.

Inspired by this key insight, we propose \textit{Denum}, a simple, general log compressor with high compression ratio and speed. 
\textit{Denum} has two main modules, \ie, \textit{Numeric Token Parsing} module and \textit{String Processing} module. \textit{Denum} first uses \textit{Numeric Token Parsing} module to extract all numeric tokens from the raw logs. Then, all the numeric tokens will be parsed and assigned distinct tags based on their patterns. For example, a numeric token conforming to the pattern \verb|<*>.<*>.<*>:<*>|, typically representing IP addresses and port numbers, will be tagged as \verb|"<I>"|. Subsequently, different processing methods are applied to numerical tokens with different tags. 
For example, timestamps, which often exhibit incremental arithmetic relationships, will be stored as the differences between successive timestamp values. Numerical tokens with different tags are stored in different files. Finally, all these generated files are compressed using a general-purpose compressor to complete the compression of numbers. For the remaining logs where numbers have been replaced with tags, we employ a \textit{String Processing} module, which includes a dictionary-index storage method, storing recurring log entries in a dictionary and using an index file to record the order of occurrences with IDs. 
\textit{Denum}'s \textit{String Processing} module can be replaced with other log compressors to adapt to specialized datasets. Extensive experiments on 16 benchmark datasets demonstrate that \textit{Denum} achieves compression ratios $8.7\%$ to $434.7\%$ higher than baselines.
In addition, \textit{Denum} achieves an average compression speed of $26.2$ MB/s across the 16 benchmark datasets, which is $2.6\times$ to $37.7\times$ faster than baselines. Moreover, \textit{Denum} is general and it can benefit existing log compressors: integrating \textit{Denum}'s \textit{Numeric Token Parsing} module into existing log compressors can yield a $11.8\%$ increase in compression ratios and a $37\%$ increase in compression speeds.

The main contributions of this paper are as follows:

\begin{itemize}
\item{We observe that many tokens in logs are numeric tokens and propose a simple yet effective idea called \textit{Numeric Token Parsing} for log compression.}
\item{We realize the idea as a log compression tool called \textit{Denum}, which contains a \textit{Numeric Token Parsing} module and a \textit{String Processing} module.}
\item{Extensive experiments are conducted on 16 widely-used log datasets and \textit{Denum} achieves the best average compression ratio and speed compared to the state-of-the-art log compressors. In addition, \textit{Denum}'s \textit{Numeric Token Parsing} module can be applied to significantly enhance the compression ratios and speeds of existing log parsers.}
\end{itemize}


\section{Preliminaries}

\subsection{General-purpose Compressor}

Employing a general-purpose compressor for log files is a straightforward approach to diminish storage expenses \cite{yao2021improving}. General-purpose compressors are designed to efficiently reduce the size of any data without specific knowledge of its characteristics. This adaptability renders general-purpose compressors highly applicable across a wide types of data.

The general-purpose compressors can be broadly categorized into four types: dictionary-based, entropy-based, sorting-based, and prediction-based.
Dictionary-based methods (e.g., LZ77 \cite{ziv1977universal} and LZ78 \cite{ziv1978compression}) construct a dynamic dictionary of data patterns encountered during the compression process. Instead of individually encoding each data element, they replace recurring data patterns with references to the dictionary. LZ77 is widely utilized in various general-purpose compression tools, such as LZMA.
Entropy-based methods (e.g., arithmetic coding \cite{rissanen1979arithmetic} and Huffman coding \cite{huffman1952method}) operate on the statistical properties of data, leveraging the concept of entropy—a quantifiable measure of unpredictability—to assign variable-length codes inversely proportional to the frequency of data patterns. Huffman coding is also widely used in compression tools. For example, the Deflate algorithm \cite{deutsch1996deflate} is a combination of the LZ77 algorithm and Huffman coding, and gzip \cite{deutsch1996deflate,gziplink} is built upon the Deflate algorithm.
Sorting-based methods reorganize data elements based on specific criteria to exploit patterns and redundancies, resulting in a more condensed representation. For example, the Burrows-Wheeler Transform (BWT) \cite{adjeroh2008burrows} used by bzip2 rearranges data to group identical elements as closely as possible. Subsequently, bzip2 applies supplementary compression techniques, including Move-To-Front (MTF) encoding \cite{manzini2001analysis}, Run-Length Encoding (RLE) \cite{golomb1966run}, and Huffman coding, to further reduce the data size.
Prediction-based methods utilize statistical models to forecast the subsequent token based on contextual information. This predictive approach effectively reduces the required bits for encoding the following token. For instance, Prediction by Partial Matching (PPM) \cite{cleary1984data} predicts forthcoming data by learning contextual patterns. The Prediction by Partial Matching with Dictionaries (PPMd) is an enhancement and extension of PPM, incorporating refinements for enhanced performance.

However, applying general-purpose compressors directly to log data is suboptimal as these compressors cannot leverage the inherent structure of log data.

\subsection{Log Compressor}

\begin{figure}[t]
	\centering
		\includegraphics[width=0.7\columnwidth]{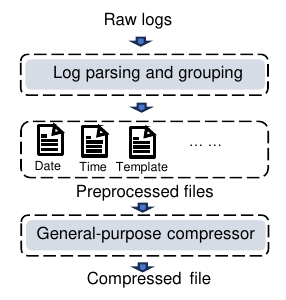}
	  \caption{The general steps of parser-based log compressor}\label{parser-based-workflow}
\end{figure}

The general steps of the parser-based log compressor are shown in Figure \ref{parser-based-workflow}, different fields are segregated and stored in separate files, referred to as preprocessed files, which are ultimately compressed using a general-purpose compressor. For example, LogZip is a classic parser-based log compressor. LogZip initially employs Drain to parse semi-structured logs into a structured format, then stores each log’s template using the dictionary-index method, meaning that all distinct log templates are stored in the form of a dictionary, and each log is stored as a template index and variables. For variables, as well as each field of the log headers, are separately stored in distinct files. Finally, a general-purpose compressor is applied to these preprocessed files. Many efforts have been dedicated to log compression, and various works augmented log parsing with an array of additional tricks to enhance performance. For instance, LogReducer \cite{wei2021feasibility} has implemented an elastic encoder specifically for the storage of numerical data and delta encoding specifically for timestamps. Building upon LogReducer, LogShrink \cite{li2024logshrink} further explores commonality and variability within log data, utilizing them for more effective compression of log data. CLP \cite{rodrigues2021clp} and LogGrep \cite{wei2023loggrep} offer efficient querying capabilities for data that has been compressed. Specifically, CLP categorizes log lines into schemas and differentiates variables into dictionary and non-dictionary categories for storage. LogGrep advances this by distinguishing between static and runtime patterns in dictionary variables, organizing them into finely segmented capsules. 

In addition to log compressors that are based on log parsers, there are some log compressors that do not utilize log parsing and achieve good results, such as LogArchieve \cite{christensen2013adaptive}, Cowic \cite{lin2015cowic}, and MLC \cite{feng2016mlc}. They adopt a unified approach in dealing with variables and templates within log data. Specifically, LogArchive leverages a similarity function along with sliding window techniques to categorize log entries into various groups, compressing these groups collectively to enhance the overall compression efficiency. Differing in its objective, Cowic prioritizes the minimization of decompression time over compression ratio by selectively decompressing only the specific log entries required, rather than the entire log file. MLC identifies and processes repetitive patterns observed in log files. Specifically, MLC employs block-level deduplication strategies to detect underlying redundancies among log entries. It then organizes these entries into clusters based on their likeness and applies delta encoding for compression.

\subsection{Number Processing in Existing Log Compressors}

\textbf{Elastic Encoder.} The elastic encoder, originating from LogReducer, was proposed for storing numbers. Since logs contain a significant number of small numerical values, storing these directly as four-byte integers leads to extensive sequences of zeros. For example, storing the number "\verb|12|" would involve three bytes of zeros like "\verb|00000000|, \verb|00000000|, \verb|00000000|, \verb|00001100|". To reduce this unnecessary overhead, the authors of LogReducer introduced the elastic encoder. This encoder uses the most significant bit (MSB) of each byte as a stop bit to achieve flexible encoding of numbers. Upon encountering a byte with an MSB of "\verb|1|", the reading of the current number ceases, and the subsequent byte immediately becomes part of the next number, eliminating the need to read four bytes for each number. For instance, the standard integer encoding of "\verb|35|" is "\verb|00000000|, \verb|00000000|, \verb|00000000|, \verb|00100011|," whereas the elastic encoding yields "\verb|10100011|." Subsequent research on log compressors, such as LogShrink, has also adopted the elastic encoder.

\textbf{Arithmetic Relationships.} Some studies have noted that general-purpose compressors fail to exploit the arithmetic relationships between numbers. For instance, LogReducer stores the delta value between two consecutive timestamps to compress them. This method can significantly reduce the size of timestamps when the target system generates logs frequently because the delta value will be much smaller than the original numbers. By using this method and elastic encoder, we can pass a much smaller number and reduce the storage overhead to the general-purpose compressor, which can improve both compression ratio and compression speed. In addition, the authors of LogReducer observed that numerical variables sometimes are correlated. For example, in an I/O trace, if the user performs sequential I/Os, the offset of the next I/O will be equal to the sum of the offset and length of the previous I/O.

\begin{figure*}[t]
	\centering
		\includegraphics[width=\textwidth]{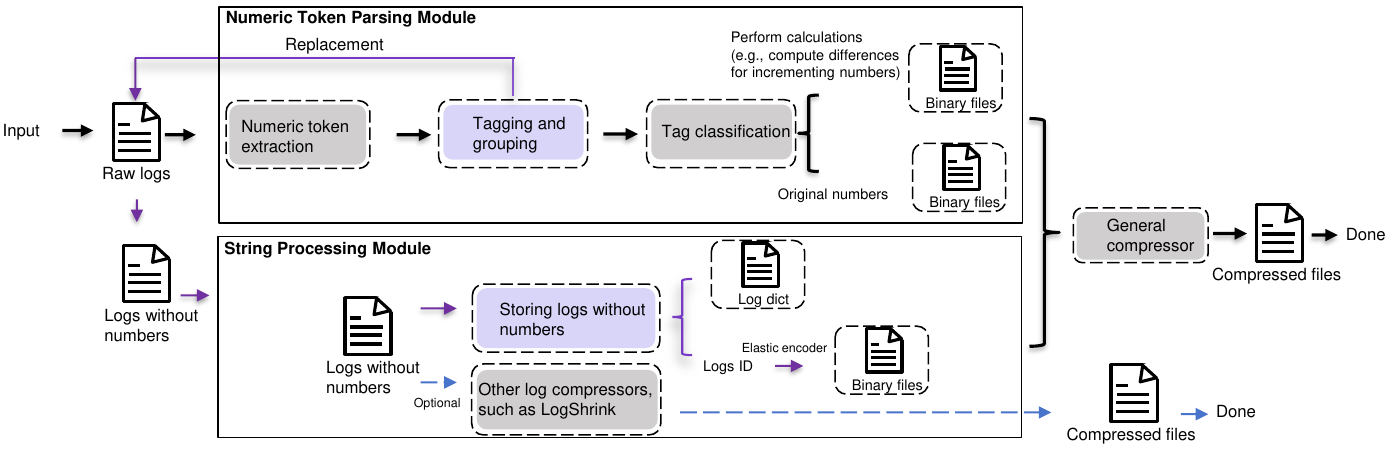}
	  \caption{The overview of \textit{Denum}}\label{Denumframework}
\end{figure*}

However, compared to pure numbers, numeric tokens can be more complex and not just numbers, it needs further parsing to obtain pure numbers for calculations. Moreover, the utilization of arithmetic relationships between numbers in logs by existing log compressors remains insufficient. Not only are timestamps printed in an incremental manner, but also various variables such as chunk IDs and counters are printed incrementally. Additionally, the current methods, which employ log parsers to group logs before processing variables, hinder the exploration of arithmetic relationships between numbers. Therefore, there is a need for a log compressor that can overcome these limitations to parse and compress the numeric tokens in logs and achieve higher compression ratios and speeds.

\section{Our Approach}

In this section, we introduce \textit{Denum}, a simple, general log compressor with high compression ratio and speed. \textit{Denum}'s framework is depicted in Figure \ref{Denumframework}. \textit{Denum} has two main modules, \ie, \textit{Numeric Tokens Parsing} and \textit{String Processing} module. The \textit{Numeric Tokens Parsing} module extracts all numeric tokens and assigns a tag to each of them. Tokens within the same tag are grouped and compressed using the same strategy. Then, the \textit{String Processing} module stores the remaining logs without numbers, using a dictionary-index method. Finally, we apply a general-purpose compressor to the files generated by the \textit{Numeric Tokens Parsing} and \textit{String Processing} modules to complete the compression process.

\subsection{Numeric Token Parsing Module}

The \textit{Numeric Token Parsing} has three steps: 1. Numeric tokens extraction, where all numeric tokens are extracted. 2. Tagging and Grouping, where all numeric tokens are assigned a tag based on their potential arithmetic relationship, and numeric tokens with the same tag are grouped together, processed into the same file. 3. Tag Classification, where different tagged groups of numeric tokens are classified into different arithmetic relationships and undergo distinct calculations.

\begin{figure}[t]
	\centering
		\includegraphics[width=0.95\columnwidth]{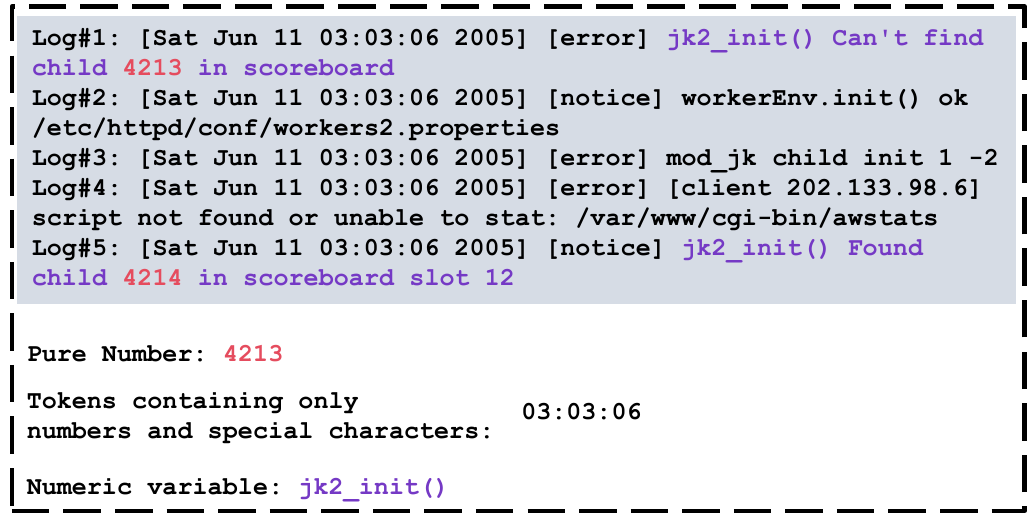}
	  \caption{Example of logs containing various types of numeric tokens}\label{logexample}
\end{figure}

\textbf{Numeric Token Extraction. }As shown in Figure \ref{logexample}, logs contain numerous numeric tokens, which we broadly categorize into three types according to their characteristics: 1. pure numbers, 2. tokens containing only numbers and special characters, and 3. numeric variables. Pure numbers can be directly extracted, while other tokens necessitate further processing to generate numbers that can be used in calculations. Pure numbers include standalone numbers, such as the code for "\verb|child|" and dates like "\verb|11|" in Figure \ref{logexample}. Tokens containing only numbers and special characters include numbers appearing with special characters, such as timestamps (e.g., "\verb|03:03:05|"), and IP addresses (e.g., "\verb|202.133.98.6|" in Figure \ref{logexample}). Numeric variables include tokens with numbers without specific meaning, often involving alphanumeric combinations, such as some MAC addresses and hash codes (e.g., "\verb|5854eb7b8b09|", "\verb|d8c05cb23ebc|"). Our method for extracting these tokens involves regular expression matching. For instance, the regular expression "\verb|R"(?<![a-zA-Z0-9])\d+(?![a-zA-Z0-9]))|" is used to extract pure numbers, while "\verb|R"((\d+).(\d+).(\d+).(\d+))"|" is employed for IP addresses. Extracted IPs like "\verb|202.133.98.6|" are not pure numeric values and cannot be directly used for calculations. Therefore, we further extract and combine the numbers into a single numeric value like "\verb|202133986|". This streamlined format allows for calculation and storage when the IP is associated with a specific arithmetic relationship.

\begin{figure}[t]
	\centering
		\includegraphics[width=1\columnwidth]{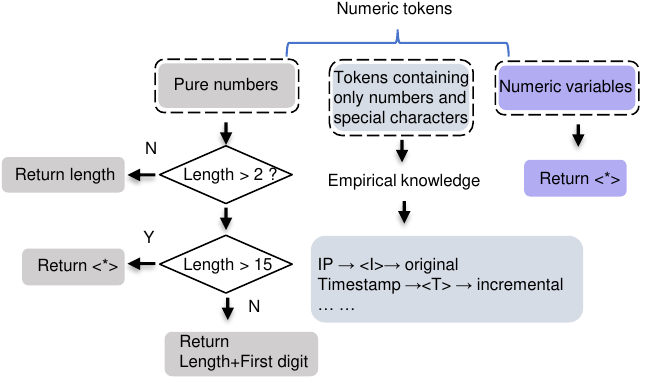}
	  \caption{Tagging numeric tokens}\label{digittags}
\end{figure}

\textbf{Tagging and Grouping.} As shown in Figure \ref{digittags}, we assign tags differently based on the type of tokens. For pure numbers, the length is an important characteristic. Numbers with a short length have small values, and even calculating the differences between these values yields results that are not significantly different from the original values. Therefore, we only calculate the differences between large values to reduce the numbers. To ensure the differences are as small as possible, our grouping considers the highest digit and the length of the numbers. Specifically, for numbers between 2 and 15 in length, we use the length as the first tag and the first digit as the second tag, \eg, "\verb|2024|" is tagged as "\verb|<4,2>|". When the number length is less than 2, the tags are their lengths, \eg, "\verb|64|" is tagged as "\verb|<2>|". For numbers longer than 15, we do not encode them with the elastic encoder due to the additional overhead required for encoding large values, thus these tokens have unified tags as \verb|"<*>"|. Finally, to avoid numbers in the tag, the generated tags will be mapped to the corresponding lowercase letter in the alphabet to generate the final tag, e.g., tag "\verb|<4,2>|" will be transformed to final tag "\verb|<db>|". For tokens containing only numbers and special characters, tag is uppercase letter and based on human empirical knowledge. For example, the pattern "\verb|<*>.<*>.<*>.<*>|" is identified by operators as an IP address pattern and is tagged as "\verb|<I>|".For numeric variables, the numbers within are meaningless, thus they are tagged as \verb|"<*>"|. Numeric tokens with the same tag will be grouped together, and all numeric tokens in the original log will be replaced by the tag.

\textbf{Tag Classification.} \textit{Denum}'s arithmetic relationship classification of tags is mainly based on empirical knowledge. For lowercase letter tags representing pure numbers, if the length of these tags is 2, indicating that the length of the numbers in the group is between 2 and 15, the consecutive numbers in this tag group will undergo a difference calculation. If the length is less than 2, we store the original values. For uppercase letter tags representing tokens containing only numbers and special characters, we determine arithmetic relationships based on empirical knowledge. For example, a timestamp denoted as "\verb|<T>|" typically increments in small decimal values in logs, so storing differences can save considerable space. For IP addresses denoted as "\verb|<I>|", there is often no arithmetic relationship between different IP addresses, so we store the original values directly. For "\verb|<*>|" representing numeric variables and long length numbers, we store the original tokens. Currently, the arithmetic relationship applied by \textit{Denum} is only incrementation. Future research could add more relationships and tags to enhance performance. 

\begin{figure}[t]
	\centering
		\includegraphics[width=0.9\columnwidth]{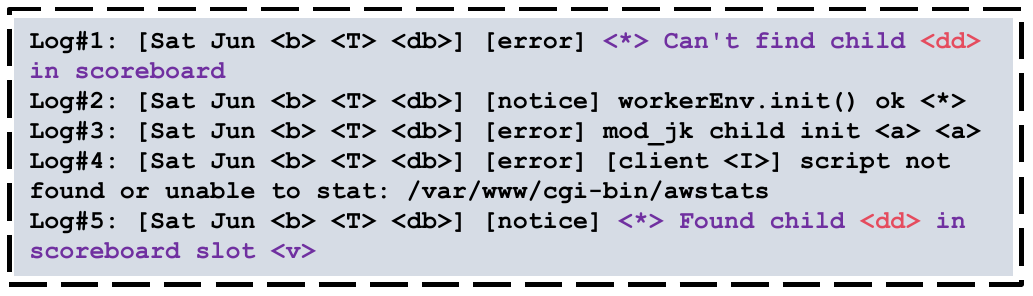}
	  \caption{Example of logs without numbers}\label{logswithoutnum}
\end{figure}

After the completion of the \textit{Numeric Token Parsing} 's operations, all numeric tokens in the logs are grouped and processed into their respective tag groups and stored as separate files, the numbers within raw logs are replaced with the corresponding tags. We refer to the logs after being replaced by tags as "logs without numbers." Figure \ref{logswithoutnum} illustrates an example of "logs without numbers," which is the result of the example log in Figure \ref{logexample} having undergone the \textit{Numeric Token Parsing}'s operations. Subsequently, the \textit{String Processing} will further compress the "logs without numbers."

\begin{figure}[t]
	\centering
		\includegraphics[width=0.9\columnwidth]{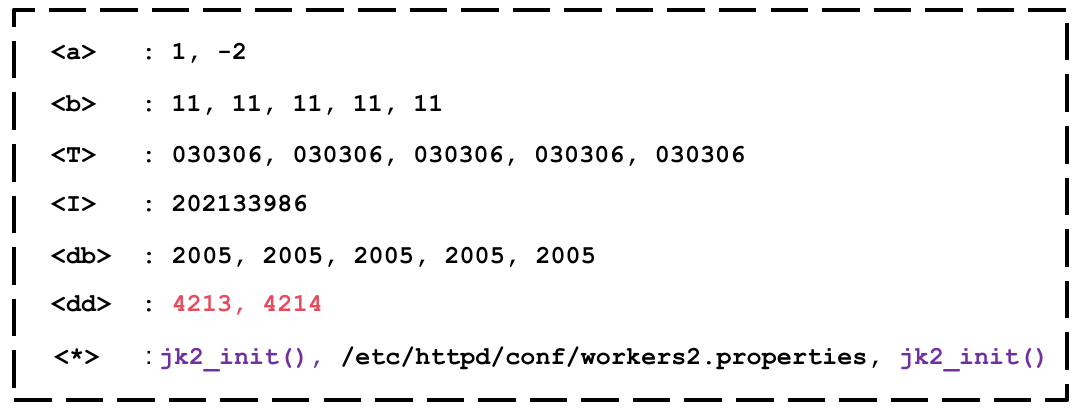}
	  \caption{Example of grouping results}\label{groupresults}
\end{figure}

\subsection{String Processing Module}

For the storage of logs without numbers, \textit{Denum} employs a \textit{String Processing} module to facilitate this process. The standard approach involves the use of a dictionary-index method, wherein these logs are converted into IDs for storage, and a dictionary is used to maintain the mapping between the IDs and the logs. Additionally, the \textit{String Processing} offers an optional feature, which allows the use of alternative log compressors. These log compressors treat logs without numbers as raw logs and use them as input for compression. This option allows other log compressors to potentially achieve higher compression ratios and speeds on specialized datasets. We designed \textit{Denum} to efficiently and effectively compress numbers, as they are prevalent in logs. This makes \textit{Denum} a foundation for log compression across various datasets. Techniques developed by other log compressors for specialized datasets can be used to further enhance compression ratios for collaborations.

Finally, a general-purpose compressor is applied to further compress log data.



\subsection{Decompressor}

\textit{Denum}, as a lossless log compressor, ensures that no information is lost when the original log files are compressed and decompressed by \textit{Denum}. The decompression process of \textit{Denum} is the reverse of the compression process. Firstly, a general compressor is first applied to decompress. Then, for the decompression of numbers, These binary files are then processed by an elastic decoder to extract the numbers. For tag groups that have undergone arithmetic operations, inverse calculations are necessary. For instance, in tag groups with an incremental relationship, the first number is successively added to subsequent differences to restore all numbers. For tag groups storing tokens containing only numbers and special characters, the numbers are recombined with special characters, such as transforming "\verb|030306|" back to "\verb|03:03:06|". For decompressing "logs without numbers," if other log compressors have been applied, their decompression algorithms are directly executed. If stored in a dictionary-index format, the binary files storing IDs are first decompressed by the elastic decoder to get log IDs, and then the dictionary is matched with the IDs in the mapping file to restore the original content. Ultimately, all numbers from the decompressed tag groups are sequentially replaced with the corresponding tags in logs without numbers. After this replacement, the decompression is finished.

\section{Evaluation}

In this section, we evaluate our approach by answering the following three research questions:
\begin{itemize}
	\item[$\bullet$] RQ1: What is the compression ratio of \textit{Denum}?
 	\item[$\bullet$] RQ2: What is the compression speed of \textit{Denum}?
	\item[$\bullet$] RQ3: Can \textit{Denum}'s \textit{Numeric Token Parsing} module improve the performance of other log compressors?
 	\item[$\bullet$] RQ4: How does each module in \textit{Denum} affect its compression ratio?
\end{itemize}

\subsection{Experimental Setting}

\textbf{Dataset.} Our experiments employ the widely accepted benchmark datasets sourced from Loghub \cite{he2020loghub}. The dataset it comprises over 378 million log entries, totaling more than 77GB, and is collected from 16 diverse systems. These systems encompass distributed computing environments like HDFS, Zookeeper, Spark, Hadoop, and OpenStack, as well as supercomputing platforms such as HPC, BGL, and Thunderbird. Furthermore, the dataset includes logs from operating systems such as Mac, Linux, and Windows, mobile operating systems like Android and HealthApp, server-based applications including Apache and OpenSSH, and standalone software applications such as Proxifier. More detailed statistics about the benchmark dataset are available in Table \ref{tab:system_log_datasets}.

\begin{table}[ht]
\centering

\caption{Description of the log datasets}
\label{tab:system_log_datasets}
\resizebox{\columnwidth}{!}{
\begin{tabular}{lccc}
\toprule
System Type & Dataset & Size & Messages \\
\midrule
\textbf{Distributed system logs} & HDFS & 1.47GB & 11,175,629 \\
 & Zookeeper & 9.95MB & 74,380 \\
 & Spark & 2.75GB & 33,236,604 \\
 & Hadoop & 48.61MB & 394,308 \\
 & OpenStack & 60.01MB & 207,820 \\
\addlinespace
\textbf{Supercomputer logs} & HPC & 32.00MB & 433,489 \\
 & BGL & 708.76MB & 4,747,963 \\
 & Thunderbird & 29.60GB & 211,212,192 \\
\addlinespace
\textbf{Operating system logs} & Mac & 16.09MB & 117,283 \\
 & Linux & 2.25MB & 25,567 \\
 & Windows & 26.09GB & 114,608,388 \\
\addlinespace
\textbf{Mobile system logs} & Android & 183MB & 1,546,686 \\
 & HealthApp & 22.44MB & 253,395 \\
\addlinespace
\textbf{Server application logs} & Apache & 4.96MB & 56,481 \\
 & OpenSSH & 70.02MB & 655,146 \\
\addlinespace
\textbf{Standalone software logs} & Proxifier & 2.42MB & 21,329 \\
\bottomrule
\end{tabular}
}
\end{table}

\textbf{Evaluation Metrics.} In our evaluation, we employ the compression ratio 
 (CR) as a metric to evaluate the effectiveness of \textit{Denum}, and compression speed (CS) to evaluate \textit{Denum}'s efficiency. These two metrics are commonly utilized in the evaluation of log compressors \cite{li2024logshrink,wei2021feasibility}. The metrics are defined as follows:

\begin{equation}
\text{Compression Ratio (CR)} = \frac{\text{Original File Size}}{\text{Compressed File Size}}
\end{equation}

\begin{equation}
\text{Compression Speed (CS)} = \frac{\text{Original File Size}}{\text{Time Taken for Compression (seconds)}}
\end{equation}

\begin{table*}[ht]
\centering
\caption{The compression ratio of \textit{Denum} and baselines}
\label{tab:compression_ratio}
\resizebox{0.8\textwidth}{!}{
\begin{tabular}{lcccccccc}
\toprule
Dataset     & gzip   & LZMA  & bzip2 & PPMd & LogZip & LogReducer & LogShrink & \textit{Denum} \\
\midrule
Android     & 7.742  & 18.857 & 12.787 & 19.370 & 25.165  & 20.776     & 21.857    & \textbf{32.494} \\
Apache      & 21.308 & 25.186 & 29.557 & 31.688 & 30.375  & 43.028     & 55.940    & \textbf{58.517} \\
BGL         & 12.927 & 17.637 & 15.461 & 18.927 & 32.655  & 38.600     & \textbf{42.385}    & 41.804 \\
Hadoop      & 20.485 & 36.095 & 32.598 & 32.110 & 35.008  & 52.830     & 60.091    & \textbf{78.546} \\
HDFS        & 10.636 & 13.559 & 14.059 & 19.155 & 16.666  & 22.634     & \textbf{27.319}    & 25.670 \\
HealthApp   & 10.957 & 13.431 & 13.843 & 15.337 & 22.632  & 31.694     & 39.072    & \textbf{44.472} \\
HPC         & 11.263 & 15.076 & 12.756 & 14.822 & 27.208  & 32.070     & 35.878    & \textbf{45.275} \\
Linux       & 11.232 & 16.677 & 14.695 & 18.508 & 23.368  & 25.213     & 29.252    & \textbf{30.449} \\
Mac         & 11.733 & 22.159 & 18.074 & 28.469 & 26.306  & 35.251     & 39.860    & \textbf{40.789} \\
OpenSSH     & 16.828 & 18.918 & 22.865 & 31.977 & 42.606  & 86.699     & \textbf{103.175}   & 101.654 \\
OpenStack   & 12.158 & 14.437 & 15.231 & 17.429 & 17.258  & 16.701     & 22.157    & \textbf{22.238} \\
Proxifier   & 15.716 & 18.982 & 23.619 & 25.489 & 21.493  & 25.501     & 27.029    & \textbf{27.288} \\
Spark       & 17.825 & 19.908 & 26.497 & 30.614 & 20.825  & 59.470     & \textbf{59.739}    & 59.470 \\
Thunderbird & 16.462 & 27.309 & 25.428 & 33.026 & —       & 49.185     & 48.434    & \textbf{63.824} \\
Windows     & 17.798 & 202.568& 67.533 & 61.083 & 310.596 & 342.975    & 456.301   & \textbf{481.350} \\
Zookeeper   & 25.979 & 27.667 & 36.156 & 38.931 & 47.373  & 94.562     & 116.981   & \textbf{135.251} \\

\bottomrule
\end{tabular}
}
\end{table*}

\textbf{Baselines.} We conducted a performance comparison of \textit{Denum} against commonly used baselines, which include LogShrink because it is state-of-the-art (SOTA) and the baselines selected by LogShrink, i.e., LogReducer and LogZip, along with several notable general-purpose compressors renowned for their effective compression ratios in log data, specifically gzip, LZMA, bzip2, and PPMd. In addition to the baselines we selected, there are several outstanding log compressors, such as LogArchive, LogBlock, and Cowic. However, LogArchive has been shown to perform significantly worse than LogReducer \cite{wei2021feasibility}. Compressors like Cowic \cite{lin2015cowic} and LogBlock \cite{yao2021improving} have specific application scenarios; Cowic focuses more on decompression speed rather than CR, while LogBlock emphasizes the compression of small log blocks. Thus, we did not select LogArchive, Cowic, and LogBlock. For experimental metrics unrelated to the experimental environment, such as compression ratios, we sourced this data directly from the original papers of each log compressor. For metrics related to the experimental environments, such as CSs, we reproduced all compressors within our experimental environment and subsequently collected their performance metrics.

\textbf{Implementation and Environment.} We performed all experiments using a Linux server, equipped with a single Intel(R) Xeon(R) Platinum 8369HB CPU @ 3.30GHz, featuring 16 cores and 32 threads due to hyper-threading. The server is provisioned with 188GB of RAM and operates on Red Hat with Linux kernel 3.10.0. For each compressor, we use 4 threads to compress the log data in parallel and sum their total time. \textit{Denum} utilizes the "tar" utility to package all preprocessed files, with the tar version being 1.30. \textit{Denum} has both Python implementation and C++ implementation. We implement the Python version of \textit{Denum} to compare with LogShrink, as it was mentioned that LogShrink's compression speed bottleneck might be due to Python's language characteristics, which are significantly slower than C++. \textit{Denum} employs the lzma compression tool as the final step. \textit{Denum}'s Python implementation is designed based on Python 3.7, and the regular expression matching was performed using the regex library, version 2022.1.18. The C++ version of \textit{Denum}, it was developed with gcc, adhering to the C++17 standard, and the regular expression library functions were implemented using PCRE2. The PCRE2\_CODE\_UNIT\_WIDTH was set to 8. All input log data is divided into 100K line blocks for subsequent processing.


\subsection{RQ1: The Compression Ratio of \textit{Denum}}

Table \ref{tab:compression_ratio} shows the CR of baselines and \textit{Denum}. Experimental results show that \textit{Denum} achieves the highest CR across 12 out of 16 benchmark datasets when compared to baselines. 
In comparison with general-purpose compressors, \textit{Denum} achieved CRs ranging from $0.736 \times$ greater than gzip (Proxifier) to $26.045 \times$ greater (Windows). When compared to LZMA, \textit{Denum}'s CRs ranged from $0.437 \times$ higher (Proxifier) to $3.888 \times$ (Zookeeper). Against bzip2, the CR varied from $0.155 \times$ greater (Proxifier) to $6.127 \times$ (Windows). Compared to PPMd, the CR varied from $0.070 \times$ greater (Proxifier) to $6.880 \times$ (Windows).
In comparison with log compressors, \textit{Denum} has a slightly lower CR on the Spark, BGL, HDFS. and OpenSSH dataset than the optimal LogShrink. Their performances are close, LogShrink achieves a CR of 42.385 on BGL, while \textit{Denum} achieves a CR of 41.804 on the same dataset, which is 98.629\% of LogShrink's performance. On the OpenSSH dataset, \textit{Denum} reaches 98.525\% of LogShrink's CR. For other datasets, \textit{Denum} achieved from $0.003 \times$ (OpenSSH) to $0.486 \times$ (Thunderbird) higher CRs. \textit{Denum}'s CR is $0.269 \times$ to $1.850 \times$ higher than LogZip. Compared to LogReducer, \textit{Denum}’s CR ranged from $0.070 \times$ greater (Proxifier) to $0.564 \times$ (Thunderbird). Notably, the authors of LogShrink reported that LogZip failed to successfully parse and compress the Thunderbird dataset within one week.

Additionally, we attempted to identify the reason of lower CR in some datasets. For example, we found that there are over a million log entries in Spark like "\verb|Update row *|", where \verb|*| represents different row numbers that vary continuously from thousands to hundreds of thousands. \textit{Denum} interprets these numbers as having an incremental relationship. Therefore, storing these logs with \textit{Denum} caused additional overhead. However, overall in Spark, \textit{Denum} still achieved a competitive CR compared to the baselines at a faster speed.

\begin{tcolorbox}
\textbf{Summary for RQ1:} \textit{Denum} outperforms existing general-purpose compressors and log compressors in terms of compression ratios on 12 out of 16 datasets. 
\end{tcolorbox}

\subsection{RQ2: The Compression Speed of \textit{Denum}}

\textit{Denum} has both Python and C++ implementations. We evaluated the CSs of both versions on 16 benchmark datasets. The experimental results are shown in Table \ref{tab:compression_speed}, \textit{Denum} was compared with the selected log compressors across 16 benchmark datasets regarding CS. Due to the reported poor CS of LogZip, we did not reproduce it. The CS of LogZip shown in Table \ref{tab:compression_speed} is from \cite{li2024logshrink}.

For the C++ implementation of \textit{Denum}, the CSs on all 16 datasets outperform other log compressors. Compared to LogShrink, the C++ implementation of \textit{Denum} was $5.300\times$ faster (HealthApp) to $44.236\times$ faster (Thunderbird). When compared to LogReducer, the C++ implementation of \textit{Denum} was $1.177\times$ faster (OpenStack) to $4.419\times$ faster (Proxifier). The average CS of the C++ implementation of \textit{Denum} across the 16 benchmark datasets was 26.204 MB/S, whereas the average compression speeds are 10.710 MB/S for LogReducer and 1.634 MB/S for LogShrink. The average speed of the C++ implementation of \textit{Denum} was $2.446\times$ that of LogReducer and $16.036\times$ that of LogShrink. For the Python implementation of \textit{Denum}, it outperforms LogShrink on all 16 benchmark datasets. The Python implementation of \textit{Denum} has an average CS of 5.515 MB/s, which is $1.421\times$ the average CS of LogShrink.

Additionally, the significant variation in \textit{Denum}'s CSs across different datasets is mainly due to the size of the original datasets. Some datasets are too small to be effectively partitioned for multithreaded parallel processing. For example, the CS of the C++ implementation of \textit{Denum} on the Linux dataset is 5.380 MB/s, and the Linux dataset contains only 25,567 entries. The block size for \textit{Denum}, LogShrink, and LogReducer is 100K. The Linux dataset is not large enough to fill even one block. Therefore, compressing the Linux dataset cannot be accelerated using multithreading.

\begin{tcolorbox}
\textbf{Summary for RQ2:} \textit{Denum} achieves much faster compression speed than baselines on all 16 datasets.
\end{tcolorbox}

\begin{table}[ht]
\centering
\caption{The compression speed (MB/S) of \textit{Denum}}
\label{tab:compression_speed}
\resizebox{\columnwidth}{!}{
\begin{tabular}{lccccc}
\toprule
Dataset & LogReducer & LogShrink & LogZip & \multicolumn{2}{c}{\textit{Denum}}  \\
\midrule
Implementation & C++ & C++ and Python & Python & C++ & Python  \\ \midrule
Android     & 14.596  & 1.300  & 0.068  & \textbf{43.027} & 7.958 \\
Apache      & 2.564  & 1.200  & 0.737  & \textbf{6.565} & 1.968\\
BGL         & 16.556 & 1.459  & 0.874  & \textbf{40.502} & 9.101 \\
Hadoop      & 7.532  & 1.907  & 0.901  & \textbf{24.647} & 6.831 \\
HDFS        & 20.093 & 1.572  & 0.701  & \textbf{46.572} & 8.101 \\
HealthApp   & 5.177  & 3.230  & 0.736  & \textbf{17.120} & 5.219 \\
HPC         & 8.465  & 2.266  & 0.644  & \textbf{26.044} & 8.339 \\
Linux       & 1.265  & 0.505  & 0.687  & \textbf{5.380} & 1.924 \\
Mac         & 5.295  & 1.059 & 0.009  & \textbf{6.746}& 2.800 \\
OpenSSH     &13.462 & 2.037  & 0.715  & \textbf{36.235} & 6.395 \\
OpenStack   & 10.389  & 1.969  & 0.537  & \textbf{12.235} & 3.915 \\
Proxifier   & 1.425  & 0.478   & 0.716  & \textbf{6.298} & 1.926 \\
Spark       & 21.875 & 1.236  & 0.550  & \textbf{51.725} & 7.077 \\
Thunderbird & 19.170 & 0.820  & ----   & \textbf{36.274} & 7.080 \\
Windows     & 18.947 & 3.618  & 1.357  & \textbf{51.350} & 8.106 \\
Zookeeper   & 4.554  & 1.497  & 0.842  & \textbf{8.549} & 1.503 \\ 
\hline
Average     & 10.710  & 1.505  & 0.694  & \textbf{26.204} & 5.515 \\ 
\bottomrule
\end{tabular}
}
\end{table}

\begin{table*}[htbp]
    \centering
    \caption{\textit{Denum}'s \textit{Numeric Token Parsing} contribution to compression ratios for various log compressors}
    \label{tab:crd}
    \begin{tabular}{lcccccc}
        \toprule
        \textbf{Dataset} & \multicolumn{2}{c}{\textbf{LogReducer}}& \textbf{Improvement}  & \multicolumn{2}{c}{\textbf{LogShrink}}& \textbf{Improvement} \\
         &Original&With \textit{Denum}&   &Original&With \textit{Denum}&  \\

        \midrule
        Android & 20.776 & 27.755 & +33.591\% & 21.857 & 27.481 & +25.731\% \\
        Apache & 43.028 & 54.111 & +25.757\% & 55.940 & 57.042 & +1.970\% \\
        BGL & 38.600 & 31.134 & -19.341\% & 42.385 & 42.439 & +0.127\% \\
        Hadoop & 52.830 & 74.036 & +40.140\% & 60.091 & 89.776 & +49.400\% \\
        HDFS & 22.634 & 23.431 & +3.521\% & 27.319 & 29.442 & +7.771\% \\
        HealthApp & 31.694 & 40.778 & +28.661\% & 39.072 & 39.510 & +1.121\% \\
        HPC & 32.070 & 40.310 & +25.693\% & 35.878 & 40.320 & +12.381\% \\
        Linux & 25.213 & 29.364 & +16.463\% & 29.252 & 33.118 & +13.216\% \\
        Mac & 35.251 & 38.616 & +9.545\% & 39.860 & 36.465 & -8.517\% \\
        OpenSSH & 86.699 & 108.634 & +25.300\% & 103.175 & 103.545 & +0.359\% \\
        OpenStack & 16.701 & 22.038 & +31.956\% & 22.157 & 21.213 & -4.261\% \\
        Proxifier & 25.501 & 25.759 & +1.011\% & 27.029 & 25.435 & -5.897\% \\
        Spark & 57.135 & 48.397 & -15.293\% & 59.739 & 59.859 & +0.201\% \\
        Thunderbird & 49.185 & 49.212 & +0.054\% & 48.434 & 50.809 & +4.904\% \\
        Windows & 342.975 & 445.872 & +30.001\% & 456.301 & 479.051 & +4.986\% \\
        Zookeeper & 94.562 & 128.401 & +35.784\% & 116.981 & 120.34 & +2.871\% \\
        \hline
        Average & 60.928 & 74.240 & +17.053\% & 74.091 & 78.490 & +5.936\% \\
        \bottomrule
    \end{tabular}
\end{table*}

\begin{table*}[htbp]
    \centering
    \caption{\textit{Denum}'s \textit{Numeric Token Parsing} contribution to compression speed for various log compressors. Due to the small size of some datasets, the difference in CR is not significant. Therefore, we alternatively record the time taken (milliseconds).}
    \label{tabcsd}
    \begin{tabular}{lcccccc}
        \toprule
        \textbf{Dataset} & \multicolumn{2}{c}{\textbf{LogReducer}}& \textbf{Improvement}  & \multicolumn{2}{c}{\textbf{LogShrink}}& \textbf{Improvement} \\
         &Original&With \textit{Denum}&   &Original&With \textit{Denum}&  \\

         \midrule
        Android & 12,740 & 11,945 & 1.067$\uparrow$ & 90,123 & 52,393 & 1.720$\uparrow$ \\
        Apache & 2,063 & 2,006 & 1.028$\uparrow$ & 4,079 & 3,550 & 1.149$\uparrow$ \\
        BGL & 39,583 & 40,937 & 0.967$\downarrow$ & 485,469 & 301,980 & 1.608$\uparrow$ \\
        Hadoop & 3,702 & 4,012 & 0.923$\downarrow$ & 24,037 & 18,301 & 1.313$\uparrow$ \\
        HDFS & 76,045 & 69,301 & 1.097$\uparrow$ & 957,170 & 543,396 & 1.761$\uparrow$ \\
        HealthApp & 4,335 & 3,182 & 1.362$\uparrow$ & 7,129 & 5,896 & 1.209$\uparrow$ \\
        HPC & 3,857 & 4,262 & 0.905$\downarrow$ & 14,120 & 12,560 & 1.124$\uparrow$ \\
        Linux & 1,800 & 2,141 & 0.841$\downarrow$ & 4,435 & 2,906 & 1.526$\uparrow$ \\
        Mac & 4,471 & 3,845 & 1.163$\uparrow$ & 15,186 & 7,863 & 1.931$\uparrow$ \\
        OpenSSH & 5,118 & 5,412 & 0.946$\downarrow$ & 34,361 & 9,694 & 3.545$\uparrow$ \\
        OpenStack & 4,595 & 6,161 & 0.746$\downarrow$ & 29,851 & 25,955 & 1.150$\uparrow$ \\
        Proxifier & 1,686 & 1,875 & 0.899$\downarrow$ & 5,063 & 2,826 & 1.792$\uparrow$ \\
        Spark & 126,759 & 134,122 & 0.945$\downarrow$ & 2,268,944 & 1,190,845 & 1.905$\uparrow$ \\
        Thunderbird & 1,621,401 & 1,639,050 & 0.989$\downarrow$ & 36,935,031 & 11,001,630 & 3.357$\uparrow$ \\
        Windows & 1,404,179 & 1,404,985 & 0.999$\downarrow$ & 7,382,040 & 4,006,757 & 1.842$\uparrow$ \\
        Zookeeper & 2,165 & 2,500 & 0.866$\downarrow$ & 6,641 & 4,575 & 1.452$\uparrow$ \\
        \hline
        Average & 207,156.188 & 208,483.500 & 0.994$\downarrow$ & 3,016,479.938 & 1,074,445.438 & 2.807$\uparrow$ \\
        \bottomrule
    \end{tabular}
\end{table*}

\subsection{RQ3: Can \textit{Denum}’s \textit{Numeric Token Parsing} module improve the performance of other log compressors?}

\textit{Denum}'s \textit{String Processing} module can be replaced with other log compressors for specialized log datasets. We evaluated the CR and total time taken by LogShrink and LogReducer using the output of \textit{Denum}'s \textit{Numeric Token Parsing} module as the original logs to compress on 16 benchmark datasets. The total time taken includes the time for \textit{Denum}'s \textit{Numeric Token Parsing} module to process the original logs and the time for these log compressors to compress the processed logs. The experimental results of CR are shown in Table \ref{tab:crd}, and Table \ref{tabcsd} presents the total time taken by log compressors.

For LogReducer, after integrating our Deunm's \textit{Numeric Token Parsing} module, the CR improved significantly on 14 out of the 16 benchmark datasets, with the highest improvement on Hadoop, reaching $40.140\%$. Before integrating Deunm's \textit{Numeric Token Parsing} module, LogReducer achieved an average CR of 60.928 across the 16 benchmark datasets. After integration, the average CR rose to 74.240 with an improvement of $17.053\%$. The most significant decrease is observed in the BGL and Spark datasets, which share a common characteristic of having a large number of irregularly numbered entries. In Spark, this appears as “\verb|Update row *|”, while in BGL, it is seen as “\verb|generating core *|.”. Denum processed these logs in a way that LogReducer and LogShrink mistakenly interpret them as new templates, such as "\verb|generating core.<cd>|, \verb|generating core.<ed>|, \verb|generating core.<ec>|." This results in additional storage overhead, such as extra template counts and more complex indexing. In contrast, by directly using Denum's simple string processing, we achieve better CR; for instance, Denum achieves a CR of 59.470 on Spark, while Denum's numeric token parsing combined with LogReducer only achieves 48.397. Adding regular expressions to capture core IDs and row IDs can further significantly enhance compression performance. LogReducer has undergone some optimizations in CS, making it the fastest among existing compression methods. After integrating Deunm's \textit{Numeric Token Parsing} module, LogReducer's CS further improved on some datasets. Experimental results show that LogReducer's CS increased on 5 out of the 16 benchmark datasets, while it slightly decreased on the others. On average, LogReducer still achieved 99.4\% of its original CS after integrating Deunm's \textit{Numeric Token Parsing} module. Overall, LogReducer, after integration with Deunm's \textit{Numeric Token Parsing} module, achieves significantly higher CRs while maintaining a comparable CS.

For LogShrink, after integrating our Deunm's \textit{Numeric Token Parsing} module, the CR improved significantly on 14 out of the 16 benchmark datasets, with the highest improvement on Hadoop, reaching $49.40\%$. Before integrating Deunm's \textit{Numeric Token Parsing} module, LogShrink achieved an average CR of 74.091 across the 16 benchmark datasets. After integration, the average CR rose to 78.490 with an improvement of $5.936\%$.
LogShrink is implemented using a combination of Python and C++, with C++ implementing the log parser. Since C++ generally executes faster than Python, the CS of LogShrink is not outstanding compared to existing C++-based log compressors. However, after integrating Deunm's \textit{Numeric Token Parsing} module, LogShrink's CS improved significantly across all datasets. Note that we did not remove any modules from LogShrink; instead, we applied our Deunm's \textit{Numeric Token Parsing} module directly before using LogShrink.
Experimental results show that LogShrink's CS improved significantly on all 16 benchmark datasets, with the highest increase reaching $3.545\times$. On average, LogShrink's CS increased to $2.807\times$ the original after integrating Deunm's \textit{Numeric Token Parsing} module. We conducted an analysis of the improvement in CS for LogShrink. LogShrink utilizes the "variability" and "commonality" in the logs to compress, where the variability in logs is often related to numbers. LogShrink employs complex pair comparisons to capture this variability, with pairs including Header-Header, Event-Header, Event-Variable, and Variable-Variable. The integration of Denum's numeric token parsing enables these number-related variabilities to be extracted easily and efficiently. As a result, LogShrink's efficiency has been enhanced. Overall, after integrating Deunm's \textit{Numeric Token Parsing} module, LogShrink achieves higher CR with nearly three times its original CS.

\begin{tcolorbox}
\textbf{Summary for RQ3:} After integrating Deunm's \textit{Numeric Token Parsing} module, LogReducer achieves a significantly higher compression ratio while maintaining a comparable compression speed. LogShrink achieves a higher compression ratio and nearly three times faster compression speed.
\end{tcolorbox}

\subsection{RQ4: How does each component in \textit{Denum} affect its compression ratio?}

\begin{figure}[t]
	\centering
		\includegraphics[width=\columnwidth]{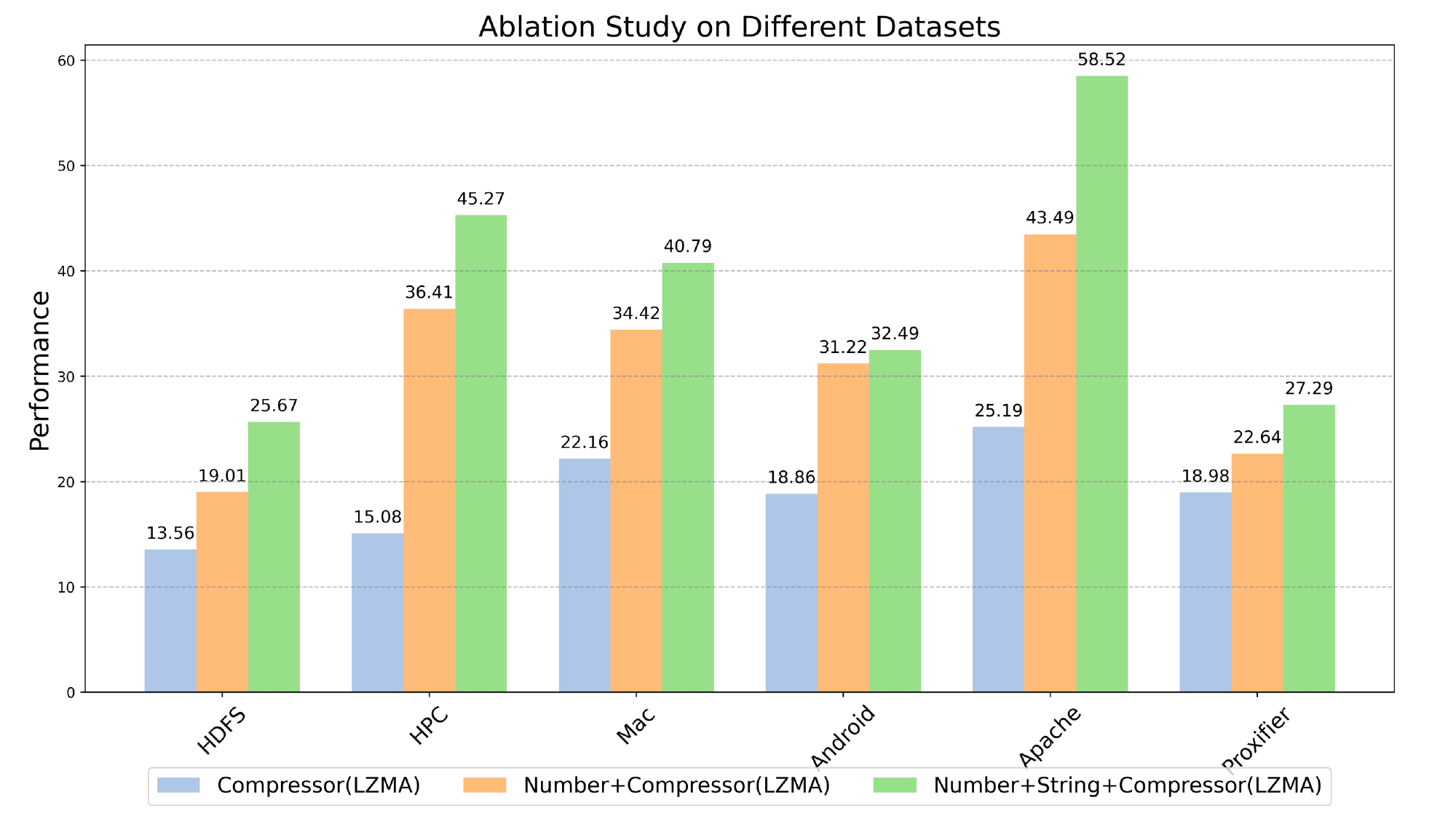}
	  \caption{Ablation study results}\label{ablation}
\end{figure}

Our ablation study is conducted on 6 datasets, which are selected from different types of systems, including HDFS (distributed system logs), HPC (supercomputer logs), Mac (operating system logs), Android (mobile system logs), Apache (server application logs), and Proxifier (standalone software logs). The main components of \textit{Denum} consist of the \textit{Numeric Token Parsing} module, \textit{String Processing} module, and a general-purpose compressor. The results of the ablation study are shown in Figure \ref{ablation}. The experimental results indicate that using only the general-purpose compressor yields poorer CRs than complete \textit{Denum}. The addition of the \textit{Numeric Token Parsing} module significantly improves the CR across all datasets. Furthermore, the inclusion of the \textit{String Processing} module results in additional improvements in CR across all datasets, demonstrating the clear contributions of each \textit{Denum} module to the overall CR. On average, the addition of the \textit{Numeric Token Parsing} module to the general-purpose compressor improves the CR by $65.800\%$, and the subsequent inclusion of the \textit{String Processing} module further enhances it by $22.800\%$.

\begin{tcolorbox}
\textbf{Summary for RQ4:} Each module of \textit{Denum} contributes to its compression ratio.
\end{tcolorbox}

\textbf{Trade-off between CR \& CS. }Commonly used general-purpose compressors such as bzip, gzip, LZMA, and PPMd have a compression level setting to trade-off between the CR and CS. For instance, when the compression level of LZMA is set to 9, the CR reaches its maximum, while the CS reaches its minimum. Similarly, for \textit{Denum}, we can achieve different compression levels of Deunm by adjusting the size of the log chunks. When we set the chunk size larger, the CS decreases, and the CR increases. Table \ref{tabcl} presents our evaluation of the CR and CS of \textit{Denum} at different compression levels across three million-scale datasets, i.e., HDFS, BGL and Spark. In our experiments, \textit{Denum}'s chunk size is set to 100K to align with LogShrink, and \textit{Denum}'s CR is slightly lower than LogShrink's in Spark, OpenSSH, HDFS and BGL under that condition. Table \ref{tabcl} shows that as the chunk size increases, \textit{Denum}'s CR gradually improves and eventually surpasses LogShrink's CR. Additionally, even with a chunk size set to 3m, \textit{Denum}'s CS remains significantly faster than LogShrink and even LogReducer.

\begin{table}[ht]
    \centering

    \caption{Compression ratio and speed for different chunk sizes (compression level) of \textit{Denum}}
    \label{tabcl}
    \resizebox{\columnwidth}{!}{
    \begin{tabular}{cccccccccc}
        \toprule
        \multirow{2}{*}{Chunk Size} & \multicolumn{3}{c}{Compression Ratio} & \multicolumn{3}{c}{Compression Speed (MB/s)} \\
        \cmidrule(lr){2-4} \cmidrule(lr){5-7}
         & BGL & HDFS & Spark & BGL & HDFS & Spark \\
        \midrule
        100K & 41.804 & 25.670 & 59.470 & 40.502 & 46.572 & 51.725 \\
        300K & 42.824 & 27.216 & 61.796 & 40.679 & 42.998 & 49.447 \\
        1M   & 43.867 & 28.986 & 63.710 & 26.309 & 35.086 & 48.096 \\
        3M   & 44.117 & 30.405 & 64.581 & 15.220 & 26.182 & 45.412 \\
        \bottomrule
    \end{tabular}
    }
\end{table}

\section{Related Work}


\textbf{Log Compressors.}  LogZip employs Drain to parse semi-structured logs into a structured format. Moreover, various works augment log parsing with an array of additional tricks to enhance performance. For instance, LogReducer utilizes an elastic encoder specifically for the storage of numerical data and delta encoding specifically for timestamps. Building upon LogReducer, LogShrink further explores commonality and variability within log data. CLP \cite{rodrigues2021clp} and LogGrep \cite{wei2023loggrep} offer efficient querying capabilities for data that has been compressed. Specifically, CLP categorizes log lines into schemas and differentiates variables into dictionary and non-dictionary categories for storage. LogGrep advances this by distinguishing between static and runtime patterns in dictionary variables, organizing them into finely segmented capsules. In addition, there are log compressors that do
not rely on log parsing, which applies unified compression strategies
to both templates and variables, such as the LogAchieve \cite{christensen2013adaptive}, Cowic \cite{lin2015cowic} and MLC \cite{feng2016mlc}.

\textbf{Number encoding methods.} Elastic encoding is proposed in LogReducer \cite{wei2021feasibility} to reduce number storage overhead. LevelDB \cite{ghemawat2014leveldb} has used variant encoding to represent numbers based on their size. Thrift \cite{slee2007thrift} has used Zigzag encoding to get more leading zero to enable efficient data serialization when communicating between processes. Compared with them, \textit{Denum}'s \textit{Numeric Token Parsing} module achieves higher compression ratios by leveraging the arithmetic relationships between numbers to reduce the values that need to be stored.


\textbf{Log Parsers.} Log parser transforms semi-structured logs into structured logs, which serves as a prerequisite for many log analysis tasks such as log-based anomaly detection and log compression. Existing log parsers can be categorized into heuristic-based \cite{he2017drain,fu2022investigating,wang2022spine}, clustering-based \cite{fu2009execution, shima2016length}, frequent pattern mining-based \cite{dai2020logram,yu2023brain}, neural network-based \cite{huo2023semparser,nedelkoski2021self,liu2022uniparser, yu2023log}, and large language model-based \cite{xu2023prompting,jiang2023llmparser, le2023evaluation, le2023log}.
Heuristic-based methods extract log templates by leveraging rules observed within logs. Frequent pattern mining-based methods consider frequently occurring patterns (e.g., words and n-grams) as template words. Clustering-based methods divide logs according to certain features. Neural network-based methods apply various neural network architectures, while large language model-based methods utilize prompts to guide large language models.

\section{Threats to Validity}

\textbf{External Validity.} \textit{Denum}'s effectiveness has been evaluated on 16 public benchmark datasets. However, these datasets might not fully capture the diversity of logs in different operational environments. Furthermore, as software evolves, the log data changes, which might make the current datasets outdated and challenge the generalizability of \textit{Denum}'s effectiveness. Therefore, future work should test \textit{Denum} on a broader range of logs to more confidently assert its external validity. In the industry, logs inevitably contain many numbers, such as timestamps. This means that \textit{Denum}'s performance on new datasets should not be too bad.


\textbf{Internal Validity.} \textit{Denum} involves using human empirical knowledge to develop regular expressions for replacing numbers, such as IP, and timestamp used in our implementations. When encountering new datasets, different regular expressions handling new number patterns except IP and time may be needed for optimal performance. This reliance on human empirical knowledge introduces variability and potential bias, affecting the superiority of \textit{Denum}’s performance across different datasets.

\section{Conclusion}

This paper proposes a simple, general log compressor with high compression ratio and speed, \textit{Denum}. \textit{Denum} has two main modules: a \textit{Numeric Token Parsing} module and a \textit{String Processing} module. The \textit{Numeric Token Parsing} module is responsible for parsing and grouping numeric tokens in the logs for subsequent processing, while the \textit{String Processing} module stores the processed logs after the numbers within are replaced. The \textit{String Processing} module can also be replaced with other log compressors for specialized log datasets. Extensive experiments show that \textit{Denum} outperforms baselines in average compression ratio on 16 benchmark datasets and offers significantly faster compression speeds. Additionally, combining \textit{Denum}'s \textit{Numeric Token Parsing} module with other log compressors can enable these compressors to achieve higher compression ratios at faster compression speeds.
\section*{DATA AVAILABILITY}
Both the C++ and Python implementations of \textit{Denum} are publicly available \cite{Denum}.

\section*{Acknowledgment}
We thank all reviewers for their insightful comments. This paper was supported by the National Natural Science Foundation of China (No. 62102340), Guangdong Basic and Applied Basic Research Foundation (No. 2024A1515010145), and Shenzhen Research Institute of Big Data Innovation Fund (No. SIF20240009).

\bibliographystyle{ACM-Reference-Format}
\bibliography{sample-base}










\end{document}